# Cryogenic Thermal Shock Effects on Optical Properties of Quantum Emitters in Hexagonal Boron Nitride

Thi Ngoc Anh Mai,[†] Sajid Ali,[‡] Md Shakhawath Hossain,[†] Chaohao Chen,[§,∥] Lei Ding, [#] Yongliang Chen,[††] Alexander S. Solntsev, [‡‡] Hongwei Mou,[#] Xiaoxue Xu,[#] Nikhil Medhekar[‡] and Toan Trong Tran[†,*]

[†]School of Electrical and Data Engineering, University of Technology Sydney, Ultimo, NSW, 2007, Australia.

[‡]School of Physics and Astronomy, Monash University, Victoria 3800, Australia.

[§]Department of Electronic Materials Engineering, Research School of Physics, The Australian National University, Canberra, Australian Capital Territory 2601, Australia.

[∥]ARC Centre of Excellence for Transformative Meta-Optical Systems (TMOS), Research School of Physics, The Australian National University, Canberra, Australian Capital Territory 2601, Australia.

[#]School of Biomedical Engineering, University of Technology Sydney, Ultimo, NSW, 2007, Australia.

[††]Department of Physics, The University of Hong Kong, Pokfulam, Hong Kong, China.

[‡‡]School of Mathematical and Physical Sciences, University of Technology Sydney, Ultimo, NSW, 2007, Australia.

*Corresponding author: trongtoan.tran@uts.edu.au

**ABSTRACT**

Solid-state quantum emitters are vital building blocks for quantum information science and quantum technology. Among various types of solid-state emitters discovered to date, color centers in hexagonal boron nitride have garnered tremendous traction in recent years thanks to their environmental robustness, high brightness and room-temperature operation. Most recently, these quantum emitters have been employed for satellite-based quantum key distribution. One of the most important requirements to qualify these emitters for space-based applications is their optical stability against cryogenic thermal shock. Such understanding has, however, remained elusive to date. Here, we report on the effects caused by such thermal shock which induces random, irreversible changes in the spectral characteristics of the quantum emitters. By employing a combination of structural characterizations and density functional calculations, we attribute the observed changes to lattice strains caused by the cryogenic temperature shock. Our study shed light on the stability of the quantum emitters under extreme conditions—similar to those countered in outer space.

Quantum information science has held the promise to revolutionize the way we communicate, resembling the way the Internet did to our society in the 20$^{th}$ century.[1-3] At the heart of quantum information science lies one of the most important pieces of quantum hardware—the quantum emitters. Such emitters can send information in the form of individual photons, also known as "the flying qubits".[4] To date, many solid-state quantum emitters have been discovered, including epitaxial quantum dots, colloidal quantum dots, single molecules, and point defects in wide bandgap materials.[1, 5-7] Among these candidates, color centers in hexagonal boron nitride (hBN) have received intense research focus thanks to their high repetition rate, robust photostability, ease of fabrication and ability to be incorporated into other van der Waals heterostructures.[8-14] The impressive traits of these quantum emitters can be attributed to the strong in-plane B-N bond that gives hBN its optical transparency, physical robustness, and chemical inertness. Most notably, however, is the robustness of the quantum emitters against extreme temperatures, up to 500°C in a vacuum as well as in harsh chemical conditions. Along with their high single photon purity and short lifetime, quantum emitters in hBN have recently been considered for satellite-based quantum communications.[15, 16] In a recent study, the quantum emitters have been subject to various types of radiation conditions including, ɣ-ray, photons and electrons, to test their suitability for space-based communication applications.[15] Surprisingly, the emitters did not show any degradation in their optical properties, making them very compelling for such applications. Nevertheless, to qualify for space-based applications, the emitters' optical attributes have to remain the same under cryogenic thermal shocks, i.e. instantaneous drop from room temperature to cryogenic temperature. This is because such cryogenic thermal shocks closely resemble conditions encountered by the emitters operating in the Earth's orbits. The impact of the cryogenic thermal shocks on hBN quantum emitters has, however, remained poorly understood to date. In this work, we perform experimental investigation on such a thermal shock effect to better understand its repercussions on the optical properties of these emitters.

## RESULTS AND DISCUSSION

We started with the sample preparation, following previous procedures published elsewhere.[17] Briefly, commercially available hexagonal boron nitride micropowder was dispersed in isopropanol (IPA) and sonicated in an ultrasound bath to exfoliate them into thinner flakes (cf. Methods). The flakes were dropcast onto a marked silicon substrate, and the substrate was annealed at 850°C under 5 mTorr of argon (50 sccm) for half an hour in a tube furnace. Such an annealing procedure has been shown to help improve the stability of the quantum emitters in hBN.[18, 19]

To examine the effect of cryogenic thermal shock on the optical properties of the hBN emitters, we designed two experiments. In the first experiment, the hBN sample was directly dropped into liquid nitrogen, causing a sudden shock in temperature experienced by the hBN flakes. As a result, the temperature dropped to liquid nitrogen temperature (77K) instantaneously, as depicted in **Figure 1a**. The sample was removed from the liquid nitrogen tank and allowed to heat up to room-temperature for further optical characterization. In the second experiment, the hBN sample was slowly cooled to liquid nitrogen temperature by using a temperature-controlled heating/cooling stage (cf. Methods) over 30 minutes, as shown in **Figure 1b**. The sample was allowed to heat up to room-temperature in a similar fashion performed as the first experiment. The second experiment is our controlled experiment as the slow cooling process closely mimics the typical cryogenic photoluminescence procedure typically conducted in optics laboratories. To prepare for both experiments, we first characterized two different groups of emitters, each on a separate marked silicon substrate. Our optical system is a home-built laser scanning confocal microscope that features a 0.7 NA air objective where excitation and emission light was focused through and collected from, respectively. A non-polarizing beam splitter (70T:30R) was used to split the excitation path from the collection path. For the excitation source, we used a 532 nm continuous-wave laser with a power of 300 μW to excite the emitters since the laser energy is well below the electronic band gap of hexagonal boron nitride, thus eliminating any unwanted excitonic emission from the hBN lattice. Specifically, we collected the emission spectra from 46 quantum emitters for the first experiment as shown in **Supporting Information Figure S1a**. During the experiments, we observed that 6 emitters bleached out after being exposed to the cryogenic thermal shock whereas the rest of the emitters were optically stable under 532 nm excitation at 300 μW power (**Supporting Information Figure S1b**). **Figure 1c** shows emission spectra taken from an exemplary emitter before (red) and after (blue) being subject to the cryogenic thermal shock. To determine the changes in the emission spectra, we fit the data with a single Lorentzian peak—an example of which can be found in the **Supporting Information Figure S2**. Surprisingly, a strong red shift of the zero-phonon line (ZPL) of ~5.5 nm occurs, from 622.46 nm to 627.95 nm, after the emitter experiences the thermal shock. Not only does the thermal shock induce the spectral shift, it also causes an increase of 4.1 nm in the linewidth, also known as the full width at half maximum (FWHM), from 6.67 nm to 10.81 nm. Such modifications in the spectral characteristics resemble the effect caused by applied strains to the hBN lattice.[12, 20] Detailed discussion will be given later in the text. On the contrary, for the second experiment, we observed no bleaching phenomenon across the 40 quantum emitters whose emission spectra were recorded in **Supporting Information Figure S3**. **Figure 1d** features two representative spectra taken from an emitter before (red) and after (blue) being exposed to the slow cooling process. Unlike the observation in the first experiment, there was no visible shift in ZPL nor change in FWHM recorded, indicating that such a slow cooling process does not affect the optical properties of the emitters per se.

To further understand the effect of cryogenic thermal shock on the emitters' optical properties, we collected statistical data of the ZPL shifts and changes in FWHM for 40 emitters in both the first and second experiment. **Figure 1e and 1f** show the ZPL shift plotted as a function of emission wavelength for the thermal shock and slow cooling experiment, respectively. For both cases, the ZPL shifts are relatively independent of the emission wavelengths. In the case of cryogenic thermal shock, the ZPL shifts are very significant, with the maximum redshift and blueshift of up to 5.5 nm and 2 nm, respectively. Conversely, for the slow cooling process, the largest redshift and blueshift are only 0.83 nm and 0.79 nm, respectively, several times smaller than that of the former case. A similar trend was observed for plots of FWHM changes versus FWHM as shown in **Figure 1g and 1h** for thermal shock and slow cooling, respectively. The magnitude of change in linewidths for the case of thermal shock is 2–3 fold larger than that of the slow cooling case. Notably, in the cryogenic thermal shock, the FWHM increases for emitters whose linewidths are around 7.5 nm or less, while the FWHM decreases for those with linewidths larger than 7.5 nm. A different tendency was observed for the case of slow cooling however. Emitters whose linewidths are smaller than 5 nm exhibit small changes in the FWHM values (< 1 nm) while those with linewidths larger than 5 nm feature significant changes in FWHM, with values clearly exceeding 1 nm. Such divided trends can be attributed to the different optical characteristics of different groups of hBN emitters.[21-23] Emitters from different groups possess dissimilar optical properties due to their unique chemical compositions and structural arrangements.[24, 25]

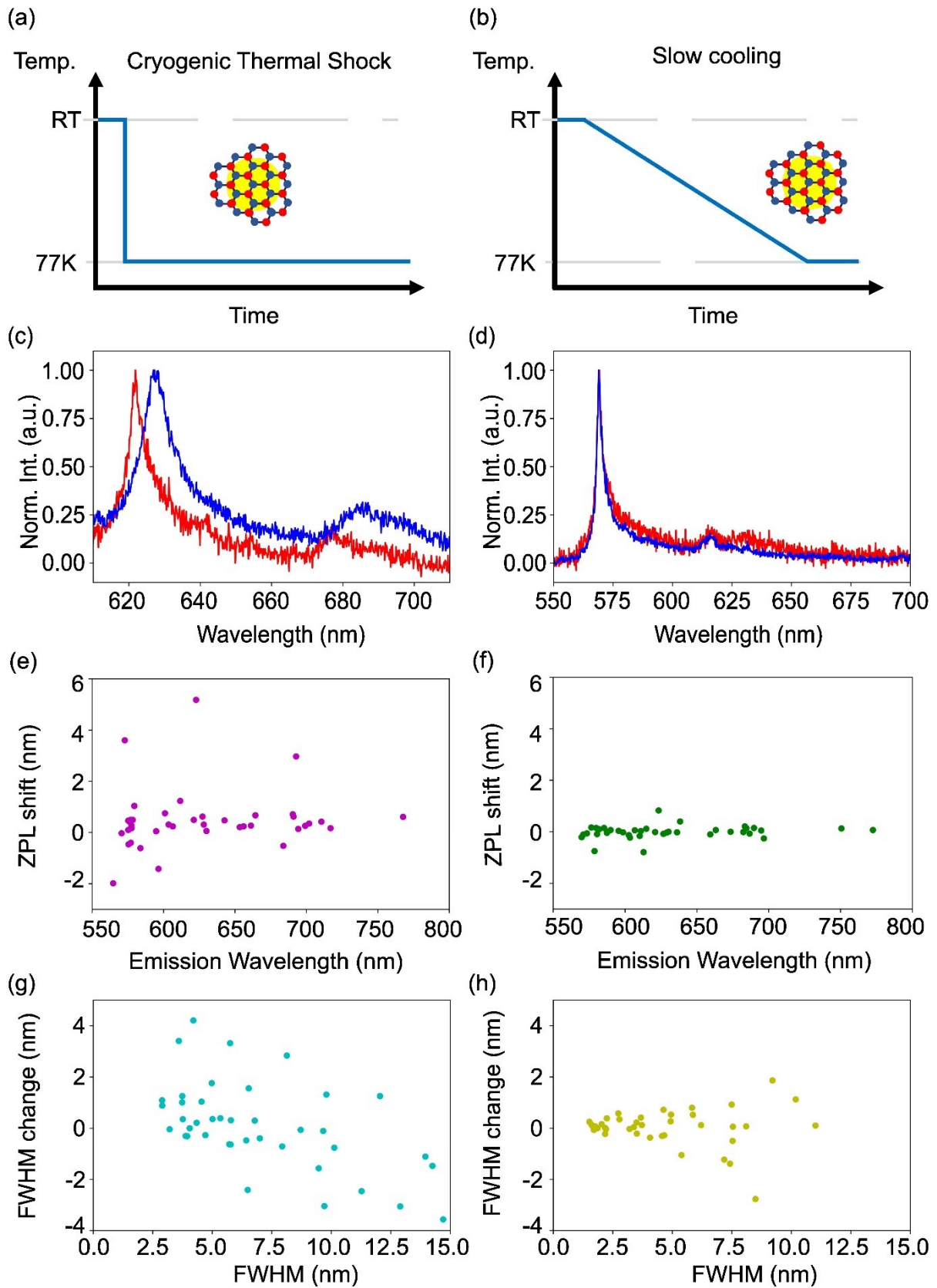

**Figure 1. Spectral changes of hBN quantum emitters before and after exposure to cryogenic thermal shock or slow cooling process. (a-b)** Schematics showing the qualitative differences in the cooling rates of cryogenic thermal shock versus slow cooling. In both

experiments, silicon substrates with hBN flakes were cooled to liquid nitrogen temperature. **(c-d)** Representative spectra of two emitters before (red) and after (blue) the thermal shock and slow cooling processes, respectively. **(e-f)** Zero-phonon line shifts plotted against emission wavelength and **(g-h)** changes in full-width at half maximum versus FWHM after the emitters are subject to thermal shock and slow cooling, respectively. Data from forty quantum emitters were chosen for (e-h). All optical characterizations in (c-h) were conducted using a 532 nm continuous-wave excitation at 300 µW at room temperature.

When the data in **Figure 1 e-h** are plotted in the form of histograms, the effect of the cryogenic thermal shock compared to that of slow cooling is clearly recognized. The thermal shock induces a ZPL shift distribution with a mean value, $\mu$, of 0.47 nm—a clear redshift—and a standard deviation, $\sigma$, of 1.18 nm **(Figure 2a)**. By contrast, the slow cooling process gives rise to a distribution with the mean value of almost zero (-0.006 nm) and a standard deviation of 0.25 nm—less than a fourth of that caused by the thermal shock **(Figure 2b)**. Such a sub-nanometer redshift in ZPLs in conjunction with a relatively large distribution width created by the cryogenic thermal shock can be a major concern for quantum applications that are wavelength-sensitive such as quantum communication or quantum teleportation—where quantum emitters need to be resonantly excited to generate the highly coherent photons.[1] In such situations, the resonant laser has to be periodically swept across the ZPL of the emitter to lock into its resonance, making the procedure arduous and time-consuming.[26] A similar trend is also observed for the FWHM of the emission spectra. **Figure 2c and 2d** entail the distribution of changes in FWHM for the case of thermal shock and slow cooling, respectively. While both the distributions are mostly symmetrical around zero, the distribution in the case of slow cooling is more than two-fold narrower (0.74 nm vs 1.72 nm) compared to that from the thermal shock case.

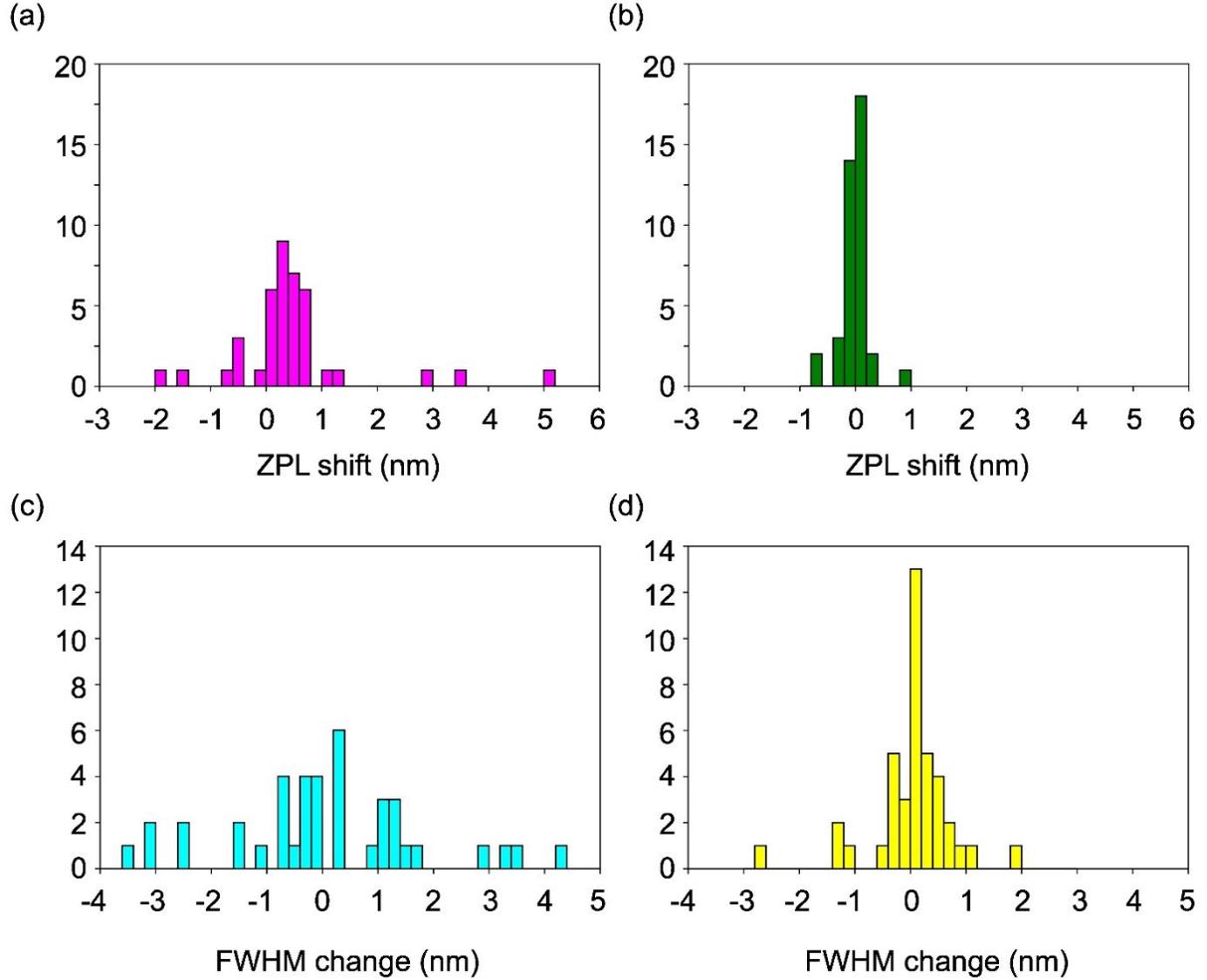

**Figure 2. Statistical analysis of modifications in spectral characteristics of the quantum emitters after the cryogenic thermal shock and slow cooling. (a-b)** Histograms of ZPL shifts after the thermal shock and slow cooling, respectively. **(c-d)** Histograms of FWHM changes after being exposed to thermal shock and slow cooling, respectively. The bin width is 0.2 nm for all the figures. Each histogram is plotted with data taken from the 40 emitters mentioned in Figure 1.

To gain more insights on the effects of thermal shock to the quantum emitters, we first looked at the morphological information from the hBN flakes before and after the thermal shock exposure. Specifically, we identified and measured three representative flakes using a standard commercial atomic force microscope (AFM) since AFM measurements are extremely sensitive in the vertical direction (z-direction). **Figure 3a** shows the topographical image featuring the three hBN flakes with various thickness, from 37.8 nm (Flake 1) to 78.6 nm (Flake 2) and 191.7 nm (Flake 3). **Figure 3b** entails the image of the same flakes after being exposed to thermal shock. Some reduction in lateral dimensions were observed across the three flakes. By analyzing the line profiles of the flakes (**Figure 3 c-e**), we noticed that there were consistent decreases in the flake thickness across all the flakes: 9.2 nm for Flake 1, 5.3 nm for Flake 2, and 81.7 nm for Flake 3. Such reduction in the flake thickness might be attributed to the sudden in-plane contraction of the hBN lattice, causing the exfoliating/peeling effect to the flakes. A reduced thickness can, in turn, give rise to variations in the dielectric environments—most notably, allowing the emitters to come closer to the trapped charges and surface states. Such proximities can induce ZPL shifts and FWHM changes via a Stark effect.

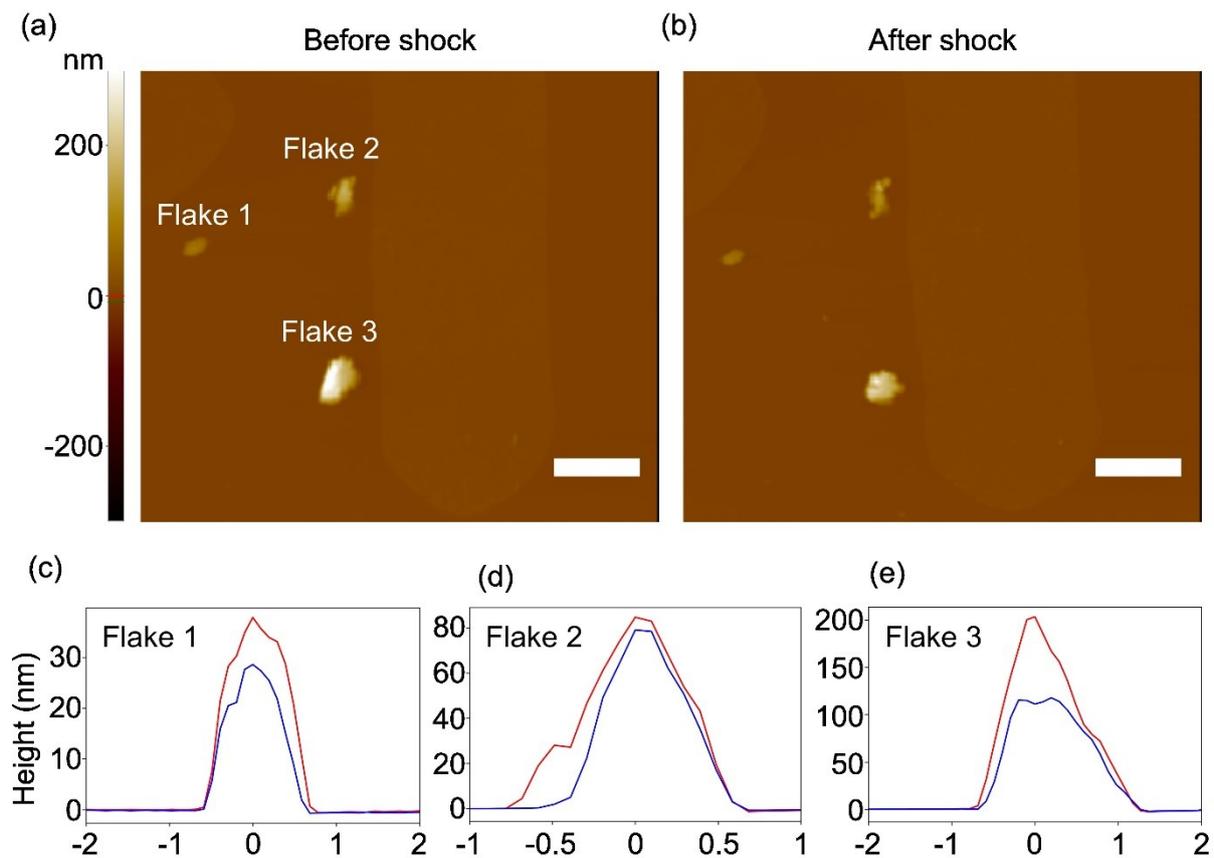

**Figure 3. AFM characterizations of hexagonal boron nitride flakes before and after the cryogenic thermal shock. (a-b)** Topographical AFM images for three hBN flakes before and after thermal shock exposure. The pixel color corresponds to the thickness of the flakes, brighter color means thicker and vice versa. The scale bars in a, b are 4 μm. **(c-e)** Line profiles of the three hBN flakes before (red line) and after (blue line) the thermal shock exposure. The faint, oval shapes in the background come from the etch-in markers used to locate the flakes in question.

While AFM offers insights on the topographical modifications of the hBN flakes after thermal shock exposure, it is not capable of providing structural information about the flakes per se. Raman spectroscopy is, on the other hand, the right tool for such characterization. By collecting the Raman spectra on a representative flake before and after the thermal shock process (cf. Methods), we can unveil any changes in the hBN crystal lattice. **Figure 4a** showcases two peaks at ~1369.2 cm$^{-1}$ obtained before (red) and after (blue) the thermal shock. The peaks can be attributed to the $E_{2g}$ in-plane phonon mode in hexagonal boron nitride. The Lorentzian-fitted spectra for both cases are, however, identical to one another, suggesting that the in-plane crystal lattice remains unchanged after the flakes are exposed to the thermal shock. To examine the quantum emission nature of the emitters after the thermal shock, we employed the second-order autocorrelation function, also known as the $g_2^{(t)}$ measurement on an exemplary emitter,[27] as shown in **Figure 4b**. Unlike the visible changes in the spectral characteristics mentioned above, we observed no significant changes in the antibunching dips (clearly below 0.5) and the antibunching times. The results indicate that there were no visible changes in the photon statistics of the emitter—a strong implication that the emitter did not undergo any compositional changes. In terms of photostability of the emitter, we notice a significant increase in fluorescence fluctuations after the emitter was subject to the thermal shock (**Figure 4c**), however. The surge in fluorescence instabilities can be attributed to the proximity to trapped charges or surface states as mentioned earlier.

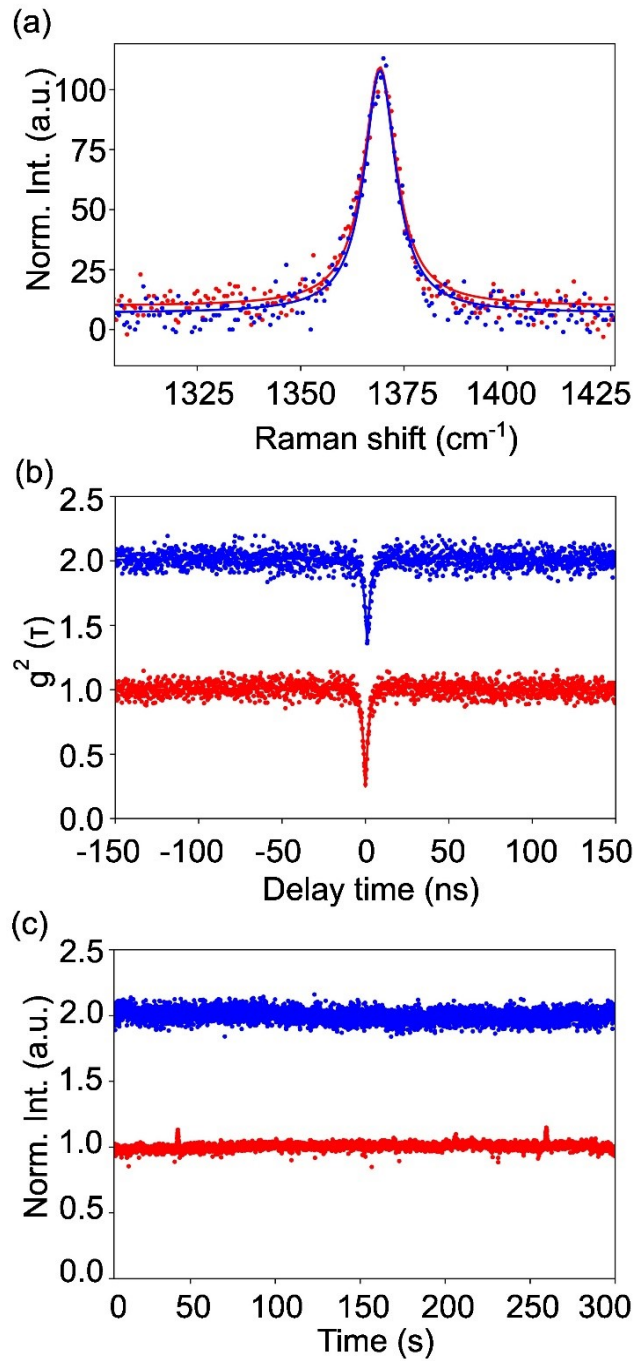

**Figure 4. Spectroscopic characterizations on hBN flakes before and after thermal shock exposure. (a)** Raman spectra, **(b)** second-order autocorrelation measurements and **(c)** photostability of a representative quantum emitter in an hBN flake before (red) and after (blue) the shock-cooling process. In (a-b), the raw and fitted data were illustrated by circle markers and solid straight lines, respectively and offset by one unit for clarity.

To obtain more in-depth information about the crystal structures of the hBN flakes subjected to the thermal shock, we resorted to the combination of transmission electron microscopy (TEM), selected area electron diffraction (SAED) and X-ray diffraction (XRD). First, we conducted the TEM imaging on a representative flake before and after the thermal shock. **Figure 5 a-b** features the images of an hBN flake in the two cases in which no visible differences can be discerned. The selected-area diffraction patterns taken from the same flake before and after the thermal shock were shown in **Figure 5 c-d**, respectively. Both reciprocal images consist of three individual diffraction patterns (red, yellow and green)—which individually belong to different sets of d-spacing of hBN in real space. It must be noted that the smaller spots within the groups of diffraction spots can be attributed to the Moire d-spacings in real space. Regardless, from our detailed analysis of these patterns, there were no significant variations in all the indexed hBN d-spacing, in good agreement with the above Raman analysis. In addition, from the TEM measurement, we can extract thickness mapping of hexagonal boron nitride flakes before and after the cryogenic thermal shock cooling process as shown in **Supporting Information Figure S4.** The mean free path (MFP) line profiles of the flake before and after the thermal shock shows a decrease in the flake thickness, which is in good agreement with the AFM measurements.

X-ray diffraction measurements were used to give further information about the out-of-plane d-spacings. **Figure 5e** depicts the XRD spectra taken for the three cases: pristine (black), after slow cooling (red) and after thermal shock (green), respectively. The peaks at the $2\theta$ angles of 26.71°, 41.10°, 43.23°, 49.11° and 53.67° can be assigned to the (002), (100), (101), (102) and (004) planes on the pristine sample, respectively. Surprisingly, except the (002) and (004) peaks, the in-plane d-spacing, which does not show any changes in the d-spacing after the thermal shock or slow cooling, the rest of the peaks that contain the out-of-plane component (the c-axis in hexagonal lattice) do show noticeable increases in the $2\theta$ angle, which translate into shrinkages in the d-spacings. Most notably, the (002) and (004) planes feature the most significant reduction in the interplanar spacing, which is in good agreement with the overall observation of the interplanar shrinkage in the hBN lattice. Such compression in the interplanar direction could be the reason for the shift in ZPLs and changes in FWHM seen previously.

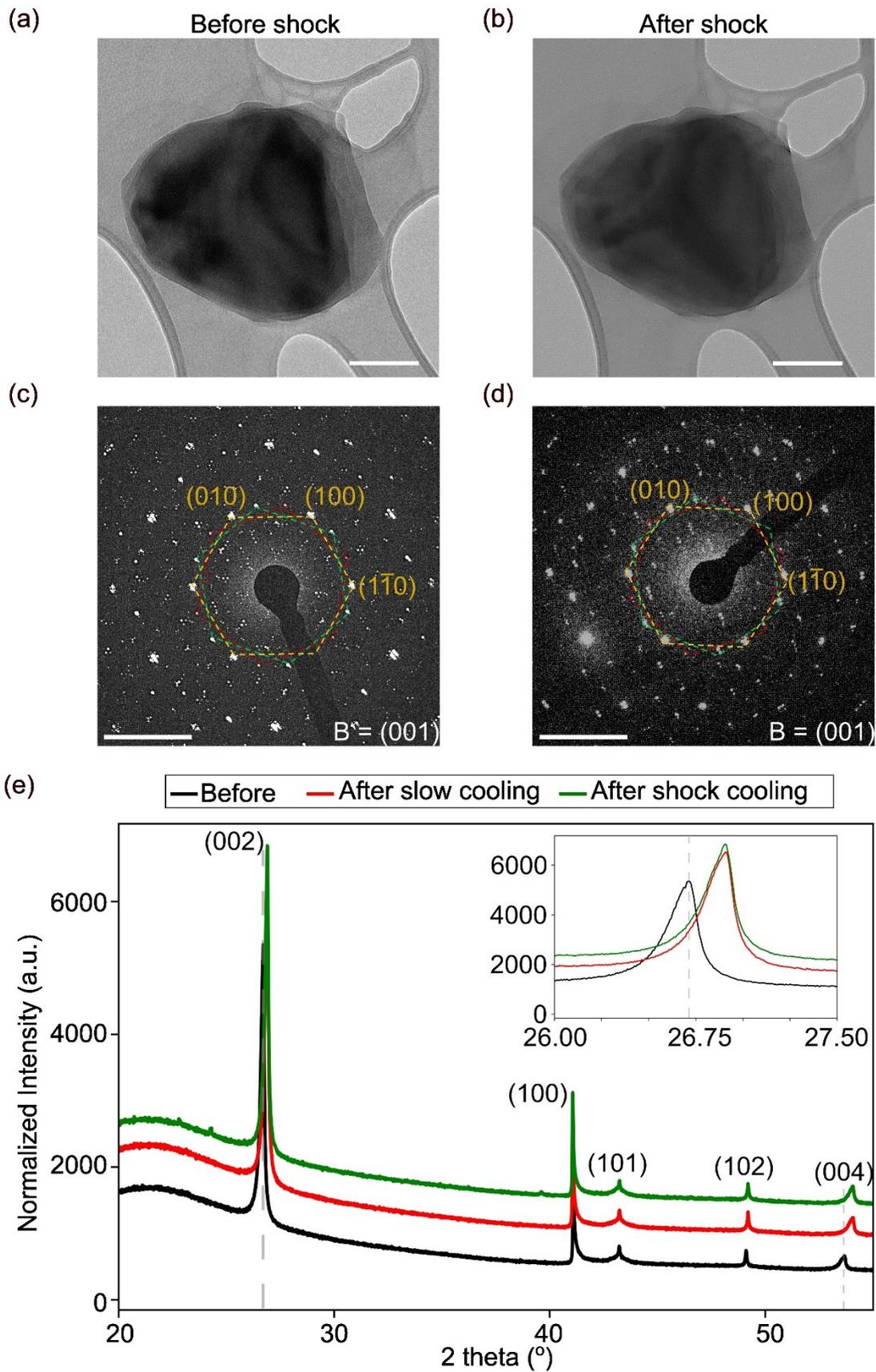

**Figure 5. Crystal structure characterization of the hBN flakes before and after cooling processes. (a-b)** Representative transmission electron microscopy (TEM) images taken before

and after the cryogenic thermal shock, respectively. The scale bars in a, b are 200 nm. **(c-d)** Representative selected area electron diffraction (SAED) images taken before and after the cryogenic thermal shock, respectively. The scale bars in c, d are 5 (1/nm). **(e)** X-ray diffraction patterns obtained from three cases: pristine, after slow cooling and after thermal shock cooling.

**DFT Calculations**

To understand the relationship between induced strain and ZPL shifts, we employed the density function theory (DFT) calculations of strain induced shift in ZPL energies and PL lineshapes. All the DFT calculations for the ground states, excited states and normal modes were performed within the VASP electronic structure code using a plane-wave wave cut-off of 800eV and a gamma point sampling of the Brillouin zone. For accurate calculation of electron spin density close to the nuclei, the projector augmented wave method (PAW) was applied together with a plane wave basis set. The defects were represented in a 9 x 9 x 1 supercell (monolayer) and atoms allowed to fully relax until the maximum force was below 0.01eV/Å. A vacuum of 15 Å was used along the vertical direction. The HSE06 exchange correlation functional was used for all calculations.

Since $(2)^2A_2 \rightarrow (1)^2A_2$ transition of the $C_2C_N$ defect (**Figure 6a**) is thought to be the source of single photon emission (SPE) from hBN.[28] We have calculated the PL lineshape for this transition and compared with the experimental luminescence spectra to confirm the origin of emission in our experiments (**Figure 6b**). Our calculated luminescence line shape plotted against the experimental spectra shows a very good agreement, given the fact that the only fitting parameter in the calculation are is gaussian smearing of 0.02eV and a rigid shift of 0.51eV to align the zero phonon lines. After confirming the origin of emission in our samples, we study the strain induced shift in the ZPL energies as shown in **Figure 6c**. We applied uniform strain within the plane of hBN and calculated the change in the ZPL energies. We find that ± 1% strain causes meV shift in the ZPL energies, consistent with experimental observations.

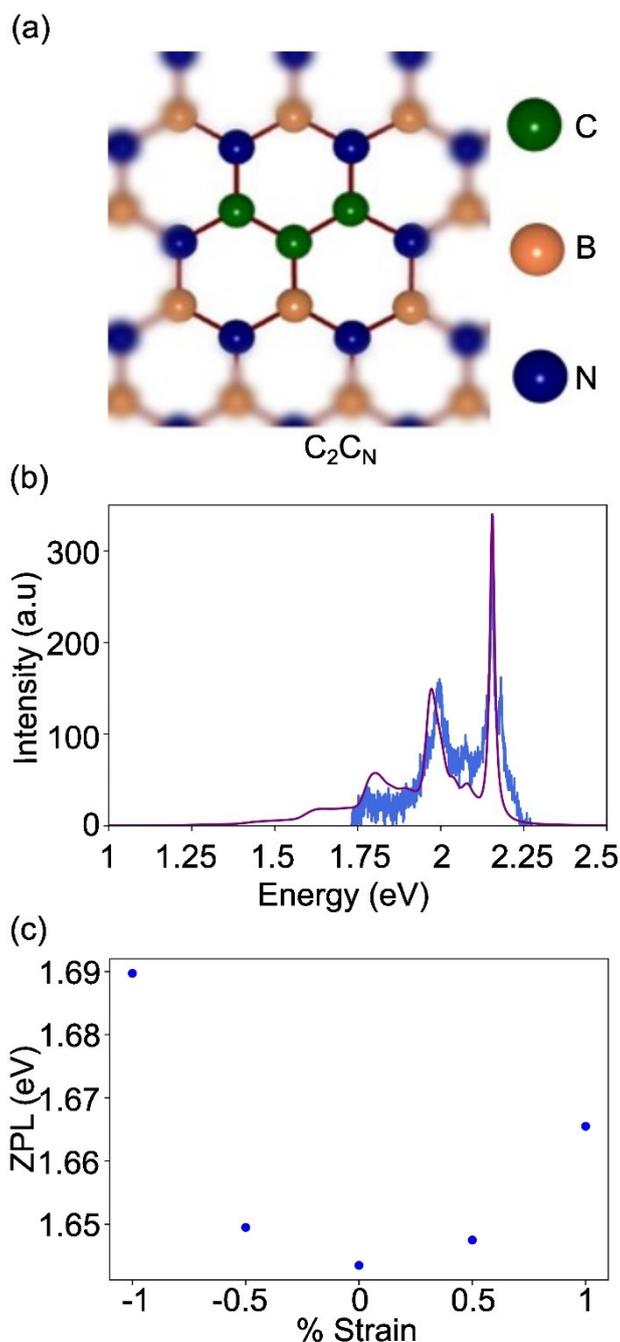

**Figure 6. Density functional theory calculations on strain-induced emission shift. (a)** $C_2C_N$ defect with $C_{2v}$ symmetry in hBN **(b)** Calculated photoluminescence spectrum compared to the experimental spectrum, the calculated spectrum with adiabatic ZPL energy has been shifted rigidly by 0.51 eV to match the high intensity peak of the experimental spectrum at 575.4 nm (2.15 eV). The blue and purple lines are the experimental data and the fit, respectively. **(c)** Strain induced shift in the ZPL energies of $(2)^2A_2 \rightarrow (1)^2A_2$ transition of the $C_2C_N$ defect.

Since the slow cooling process has only minor effects on the optical properties of the emitters, we mainly focus to the effect caused by the thermal shock process. For practical implementations, especially in the case of space-based applications, quantum emitters need to be spectrally stable against multiple cryogenic thermal shock cycles. As such, we designed an experiment in which the quantum emitters were exposed to three thermal shock cycles to gain more insights on the effect of repeated exposure to the thermal shock on the quantum emitters. After each cycle, we optically characterized the same emitters and recorded their spectral data, similar to that in the earlier experiments. To be as consistent with the previous experiments as we can, we conducted the second and third thermal shock cycles on the same sample that was exposed to the first thermal shock cycle (in **Figure 1 c, e** and **g**). As mentioned earlier, we started the experiment with 46 emitters. After each cycle, a small portion of emitters bleached out. Specifically, after the first, second and third thermal cycle, the numbers of emitters that bleached out were 6, 7 and 3, respectively. After the third cycle, only 30 emitters remained optically stable while the other 16 emitters were optically inactive. This means that ~35% of emitters bleached out after the three thermal cycles—implying the detrimental effect of the consecutive cryogenic thermal shocks on the optical properties of the emitters. The remaining thirty emitters were identified and characterized, and their spectra were fitted using a single Lorentzian peak. To compare the ZPL shifts and FWHM changes after each thermal shock cycle, we subtracted the fitted values obtained from each thermal shock cycle to that from the original emission before the thermal shock, $\gamma_{r,i} = \gamma_i - \gamma_0$, where $\gamma_{r,i}$ is the subtracted value, $\gamma_i$ is the fitted value after thermal shock cycle i, and $\gamma_0$ is the fitted value before the first thermal shock. **Figure 7a and 7b** show the ZPL shifts plotted in category and box plots, respectively, for three consecutive thermal shock cycles. It is clear that the second and third thermal cycles cause significantly more ZPL shifts, on average, than that from the first cycle, with the median values of 1.09 and 0.90 nm, respectively, compared to 0.31 nm from the first cycle. Moreover, compared to the first and second cycles, the third cycle induces more than two-fold increase in the distribution of ZPL shifts—1.12 nm versus 0.39 and 0.43, respectively. Such an increase in the distribution width is also observed for the FWHM changes as shown in **Figure 7c and 7d**.

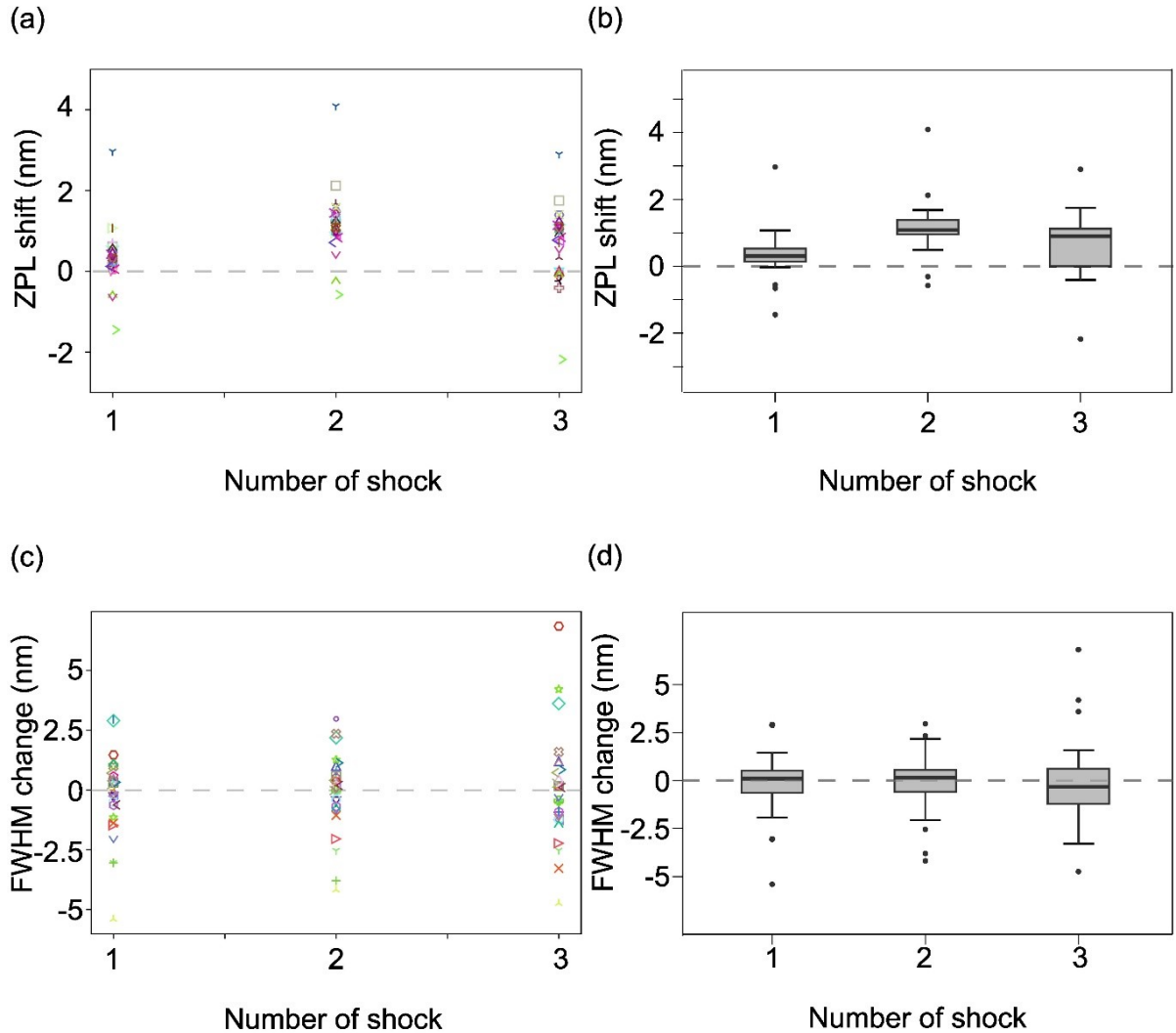

**Figure 7. Modifications in optical characteristics of hBN quantum emitters upon three consecutive cryogenic thermal shock cycles.** The ZPL shifts after three consecutive thermal shock cycles plotted as category (**a**) and box plot (**b**), respectively. The FWHM changes after three consecutive thermal shock cycles plotted as category (**c**) and box plot (**d**), respectively. A total number of thirty emitters were considered for this experiment. All data were calculated by subtracting the fitted values after each thermal shock to the fitted values before the thermal shock. The grey dash line at zero was added as the reference to improve visualization.

## CONCLUSIONS

In this study, we have presented a comprehensive analysis of the cryogenic thermal shock effects on the optical properties of quantum emitters embedded in hBN. Our experimental data, supported by structural characterizations and density functional theory calculations, demonstrate that cryogenic thermal shocks induce spectral shifts in these emitters. This is a result of lattice strains, which lead to random and irreversible alterations in their emission characteristics. Notably, these findings underscore the sensitivity of hBN quantum emitters to extreme temperature fluctuations, despite their otherwise robust performance under high-temperature and chemically harsh environments. The resilience of these quantum emitters to radiation further accentuates their potential for space-based applications, particularly in quantum communication technologies.

Looking ahead, the implications of our research are twofold. Firstly, there is a clear imperative to further explore materials engineering strategies that could enhance the thermal shock resistance of hBN quantum emitters. This could involve the development of novel fabrication techniques or the investigation of composite materials that could mitigate lattice strain effects. Secondly, our findings pave the way for new experimental protocols that specifically simulate the harsh conditions of outer space, providing a more rigorous testing ground for quantum emitters in future quantum communication networks. As the quest for quantum-enabled technologies progresses, the insights provided by our research will be invaluable in navigating the challenges posed by the extreme conditions of space, ensuring that the promise of quantum communication can be realized fully and reliably.

## METHODS

### Sample preparation

The hexagonal boron nitride (hBN) flakes used in this work were prepared by solvent exfoliation method.[17] 10 mg hBN powder of Plasma Chem was added into 20 ml isopropanol (IPA), followed by a probe ultrasonication for 10 minutes for exfoliation, and diluted 100 times in IPA. hBN diluted solution was dropped onto a 1 cm x 1 cm silicon substrate. Then it was heated at 80°C on a hotplate for 10 minutes to dry. Finally, the substrate was annealed at 850°C in a tube furnace (Lindberg Blue Mini-mite) with 50 sccm argon flow rate for 30 minutes to activate and stabilize the hBN flakes. The sample was stored in a vacuum bag to avoid other contamination for further experiments.

### Thermal shock experiments

### Cryogenic thermal shock process

The silicon substrate with the hBN flakes was dipped directly in a 30L liquid nitrogen (LN2) dewar using a bucket/canister. After waiting for 3 minutes the sample was taken out immediately to room temperature and finally dried naturally for 10 minutes.

**Slow-cooling process**

The silicon substrate with the hBN flakes was placed in a thermal vacuum heating/cooling stage (Microoptik-MHCS-622) assembled with a high-performance temperature controller (Microoptik- MDTC600) with temperature accuracy of 0.1°C, a liquid nitrogen pump (LN2-SYS) and a liquid nitrogen (LN2) dewar. The sample was glued on the stage with a silver conductive grease to improve the heat transfer process. The sample in the stage was cooled down from room temperature to -190°C gradually for 30 minutes with a cooling rate of 6°C/minute, stayed at -190°C for 10 minutes, and then heated up to room temperature slowly in 6 hours.

**Optical characterization**

Photoluminescence and Raman spectra of all single-photon emitters in hexagonal boron nitride samples before and after cryogenic thermal shock and slow cooling process were collected by an Andor spectrometer (SR-500i-A) equipped with a CCD camera in a lab-made confocal microscope set up.[29] The sample was excited with a 532 nm continuous wave (CW) laser (Cobolt Samba 532 nm) with an excitation power of 300 μW. A scanning mirror (Newport, SFM-CD300B) was used to navigate the laser spot position in the sample plane to generate the confocal map. A 4F system and an objective that has a numerical aperture (NA) of 0.7 (Thorlabs, MY100X-806, 100×) were used to focus the laser spot on the surface of the sample which was put on a three-dimensional micro-positioner stage. A tunable neutral density (ND) filter was placed on the excitation path to adjust the excitation power of the laser. The collection and excitation arms were separated by a cube beam splitter (30R/70T; Thorlabs). A long pass filter (Semrock, LP 561) was placed in the collection path to completely suppress the 532 nm excitation signal. The emission signal from the collection arm was divided into two pathways by a plate beam splitter (50R/50T; Thorlabs), coupled into two graded-index multimode fibers (Thorlabs, GIF625). One path was connected to a single-photon avalanche photodiode (SPAPD) (Excelitas Technologies, SPCM-AQRH 14-FC) to count all the incoming photons and the other was connected to the spectrometer to visualize the spectra from each single photon emitter. LabVIEW software was used to control all the hardware, analyze the photon rates, and generate the confocal map based on the signal collected in SPAPD. Andor Solis software was used to analyze the spectra collected in the spectrometer. For each single-photon emitter, zero phonon line and phonon sideband were fitted with the Lorentzian function to extract both the peak position and the full-width half maximum (FWHM) of the peak. An example of a fitted spectrum is shown in **Supporting Information Figure S2**. The second-order autocorrelation measurements were conducted with the Hanbury Brown-Twiss setup equipped with two SPAPDs and a time-correlated single photon counting module (Time Tagger 20).[30]

## Structural characterization

### Atomic force microscopy

Atomic force microscopy (AFM) images were taken using the Park XE7 under ambient conditions. A standard silicon probe (Nanoworld, NCH, spring constant of 42 N/m and a natural frequency of 320 kHz) was used. The images were flattened and corrected using XEI software (Park Systems).

### Transmission electron microscopy

A JEOL2200FS TEM was used to characterise the morphology and the selected area electron diffraction (SAED) patterns of a selected hBN flakes on copper grid. The copper grid was taken out and repeated the cryogenic thermal shock process. The copper grid was put back to the TEM for the morphology and SAED pattern characterization on the same hBN flake. The accelerating voltage was 200 kV.

### X-ray diffraction

X-ray Diffraction patterns for the same batch hBN powder were obtained using a Bruker D8 Discover A25 X-ray diffractometer with Cu K1 radiation (40 kV, 40 mA, λ=0.15406 nm). The hBN powder was dispersed in ethanol and then drops-casted on a Si wafer, then the XRD measurement was proceeded at room temperature. Then the hBN powder on the Si wafer was placed into LN directly and kept for 3 mins. After it was taken out, XRD measurement was carried out again at the same conditions. The scan step was 0.02 deg/min and scanning time was 0.4 s/step.

### Density functional theory

The density functional theory (DFT) calculations for the ground states, excited states and normal modes were conducted using the VASP electronic structure code with a plane-wave wave cut-off of 800 and a gamma point sampling of the Brillouin zone. For accurate calculation of electron spin density close to the nuclei, the projector augmented wave method (PAW) was applied together with a plane wave basis set. The defects were represented in a 9 x 9 x 1 supercell (monolayer) and allowed to fully relax until the maximum force was below 0.01eV/Å. A vacuum of 15 Å was used along the vertical direction. The HSE06 exchange correlation functional was used for all calculations.

## ASSOCIATED CONTENT

**Data Availability Statement**

The datasets generated during and/or analysed during the current study are available from the corresponding author on reasonable request.

## AUTHOR INFORMATION

**Author contribution**

T. N. A. M. and T. T. T conceived the idea of the project. T. N. A. M. and T. T. T built the optical system and its software. T. N. A. M. fabricated the quantum emitters in hBN and performed all the optical characterization. X. X. and H. M. conducted the TEM, SAED and XRD experiments and analyzed the data. S. A. and N. M. carried out the DFT calculations for strain effect on hBN quantum emitters. T. T. T supervised the project. All authors discussed the results and commented on the manuscript.

**Acknowledgement**

T. N. A. M. is grateful to UTS and VNU for the financial support to this research through the Joint Technology and Innovation Research Centre (JTIRC) scheme. The authors thank the UTS node of Optofab ANFF for the assistance with nanofabrication. The authors acknowledge the technical and scientific assistance of Sydney Microscopy & Microanalysis, the University of Sydney node of Microscopy Australia.

**Funding Sources**

T. T. T. acknowledges the Australian Research Council (DE220100487) for the financial support.

**Notes**

The authors declare no competing financial interest.

**References**

(1) Wolfowicz, G.; Heremans, F. J.; Anderson, C. P.; Kanai, S.; Seo, H.; Gali, A.; Galli, G.; Awschalom, D. D. Quantum guidelines for solid-state spin defects. *Nature Reviews Materials* **2021**, *6*, 906-925.
(2) Awschalom, D. D.; Hanson, R.; Wrachtrup, J.; Zhou, B. B. Quantum technologies with optically interfaced solid-state spins. *Nature Photonics* **2018**, *12*, 516-527.
(3) Atatüre, M.; Englund, D.; Vamivakas, N.; Lee, S.-Y.; Wrachtrup, J. Material platforms for spin-based photonic quantum technologies. *Nature Reviews Materials* **2018**, *3*, 38-51.
(4) Wei, S.-H.; Jing, B.; Zhang, X.-Y.; Liao, J.-Y.; Yuan, C.-Z.; Fan, B.-Y.; Lyu, C.; Zhou, D.-L.; Wang, Y.; Deng, G.-W.; et al. Towards Real-World Quantum Networks: A Review. *Laser & Photonics Reviews* **2022**, *16*, 2100219.
(5) Aharonovich, I.; Englund, D.; Toth, M. Solid-state single-photon emitters. *Nature Photonics* **2016**, *10*, 631-641.
(6) Uppu, R.; Midolo, L.; Zhou, X.; Carolan, J.; Lodahl, P. Quantum-dot-based deterministic photon–emitter interfaces for scalable photonic quantum technology. *Nature Nanotechnology* **2021**, *16*, 1308-1317.
(7) Montblanch, A. R. P.; Barbone, M.; Aharonovich, I.; Atatüre, M.; Ferrari, A. C. Layered materials as a platform for quantum technologies. *Nature Nanotechnology* **2023**, *18*, 555-571.
(8) Tran, T. T.; Bray, K.; Ford, M. J.; Toth, M.; Aharonovich, I. Quantum emission from hexagonal boron nitride monolayers. *Nature Nanotechnology* **2016**, *11*, 37-41.
(9) Martínez, L. J.; Pelini, T.; Waselowski, V.; Maze, J. R.; Gil, B.; Cassabois, G.; Jacques, V. Efficient single photon emission from a high-purity hexagonal boron nitride crystal. *Physical Review B* **2016**, *94*, 121405.
(10) Tran, T. T.; Elbadawi, C.; Totonjian, D.; Lobo, C. J.; Grosso, G.; Moon, H.; Englund, D. R.; Ford, M. J.; Aharonovich, I.; Toth, M. Robust Multicolor Single Photon Emission from Point Defects in Hexagonal Boron Nitride. *ACS Nano* **2016**, *10*, 7331-7338.
(11) Exarhos, A. L.; Hopper, D. A.; Grote, R. R.; Alkauskas, A.; Bassett, L. C. Optical Signatures of Quantum Emitters in Suspended Hexagonal Boron Nitride. *ACS Nano* **2017**, *11*, 3328-3336.
(12) Grosso, G.; Moon, H.; Lienhard, B.; Ali, S.; Efetov, D. K.; Furchi, M. M.; Jarillo-Herrero, P.; Ford, M. J.; Aharonovich, I.; Englund, D. Tunable and high-purity room temperature single-photon emission from atomic defects in hexagonal boron nitride. *Nature Communications* **2017**, *8*, 705.
(13) Jungwirth, N. R.; Fuchs, G. D. Optical Absorption and Emission Mechanisms of Single Defects in Hexagonal Boron Nitride. *Physical Review Letters* **2017**, *119*, 057401.
(14) Aharonovich, I.; Tetienne, J.-P.; Toth, M. Quantum Emitters in Hexagonal Boron Nitride. *Nano Letters* **2022**, *22*, 9227-9235.
(15) Vogl, T.; Sripathy, K.; Sharma, A.; Reddy, P.; Sullivan, J.; Machacek, J. R.; Zhang, L.; Karouta, F.; Buchler, B. C.; Doherty, M. W.; et al. Radiation tolerance of two-dimensional material-based devices for space applications. *Nature Communications* **2019**, *10*, 1202.

(16) Ahmadi, N.; Schwertfeger, S.; Werner, P.; Wiese, L.; Lester, J.; Da Ros, E.; Krause, J.; Ritter, S.; Abasifard, M.; Cholsuk, C. QUICK $^3$--Design of a satellite-based quantum light source for quantum communication and extended physical theory tests in space. *arXiv preprint arXiv:2301.11177* **2023**.
(17) Chen, Y.; Westerhausen, M. T.; Li, C.; White, S.; Bradac, C.; Bendavid, A.; Toth, M.; Aharonovich, I.; Tran, T. T. Solvent-Exfoliated Hexagonal Boron Nitride Nanoflakes for Quantum Emitters. *ACS Applied Nano Materials* **2021**, *4*, 10449-10457.
(18) Vogl, T.; Campbell, G.; Buchler, B. C.; Lu, Y.; Lam, P. K. Fabrication and Deterministic Transfer of High-Quality Quantum Emitters in Hexagonal Boron Nitride. *ACS Photonics* **2018**, *5*, 2305-2312.
(19) Stewart, J. C.; Fan, Y.; Danial, J. S. H.; Goetz, A.; Prasad, A. S.; Burton, O. J.; Alexander-Webber, J. A.; Lee, S. F.; Skoff, S. M.; Babenko, V.; et al. Quantum Emitter Localization in Layer-Engineered Hexagonal Boron Nitride. *ACS Nano* **2021**, *15*, 13591-13603.
(20) Mendelson, N.; Doherty, M.; Toth, M.; Aharonovich, I.; Tran, T. T. Strain-Induced Modification of the Optical Characteristics of Quantum Emitters in Hexagonal Boron Nitride. *Advanced Materials* **2020**, *32*, 1908316.
(21) Mendelson, N.; Chugh, D.; Reimers, J. R.; Cheng, T. S.; Gottscholl, A.; Long, H.; Mellor, C. J.; Zettl, A.; Dyakonov, V.; Beton, P. H.; et al. Identifying carbon as the source of visible single-photon emission from hexagonal boron nitride. *Nature Materials* **2021**, *20*, 321-328.
(22) Koperski, M.; Nogajewski, K.; Potemski, M. Single photon emitters in boron nitride: More than a supplementary material. *Optics Communications* **2018**, *411*, 158-165.
(23) Shaik, A. B. D.-a.-j.-w.-i.; Palla, P. Optical quantum technologies with hexagonal boron nitride single photon sources. *Scientific Reports* **2021**, *11*, 12285.
(24) Hayee, F.; Yu, L.; Zhang, J. L.; Ciccarino, C. J.; Nguyen, M.; Marshall, A. F.; Aharonovich, I.; Vučković, J.; Narang, P.; Heinz, T. F.; et al. Revealing multiple classes of stable quantum emitters in hexagonal boron nitride with correlated optical and electron microscopy. *Nature Materials* **2020**, *19*, 534-539.
(25) Shotan, Z.; Jayakumar, H.; Considine, C. R.; Mackoit, M.; Fedder, H.; Wrachtrup, J.; Alkauskas, A.; Doherty, M. W.; Menon, V. M.; Meriles, C. A. Photoinduced Modification of Single-Photon Emitters in Hexagonal Boron Nitride. *ACS Photonics* **2016**, *3*, 2490-2496.
(26) Dietrich, A.; Bürk, M.; Steiger, E. S.; Antoniuk, L.; Tran, T. T.; Nguyen, M.; Aharonovich, I.; Jelezko, F.; Kubanek, A. Observation of Fourier transform limited lines in hexagonal boron nitride. *Physical Review B* **2018**, *98*, 081414.
(27) Fleury, L.; Segura, J. M.; Zumofen, G.; Hecht, B.; Wild, U. P. Nonclassical Photon Statistics in Single-Molecule Fluorescence at Room Temperature. *Physical Review Letters* **2000**, *84*, 1148-1151.
(28) Fischer, M.; Sajid, A.; Iles-Smith, J.; Hötger, A.; Miakota, D. I.; Svendsen, M. K.; Kastl, C.; Canulescu, S.; Xiao, S.; Wubs, M.; et al. Combining experiments on luminescent centres in hexagonal boron nitride with the polaron model and ab initio methods towards the identification of their microscopic origin. *Nanoscale* **2023**, *15*, 14215-14226.
(29) Chen, C.; Ding, L.; Liu, B.; Du, Z.; Liu, Y.; Di, X.; Shan, X.; Lin, C.; Zhang, M.; Xu, X.; et al. Exploiting Dynamic Nonlinearity in Upconversion Nanoparticles for Super-Resolution Imaging. *Nano Letters* **2022**, *22*, 7136-7143.
(30) Chen, Y.; White, S.; Ekimov, E. A.; Bradac, C.; Toth, M.; Aharonovich, I.; Tran, T. T. Ultralow-Power Cryogenic Thermometry Based on Optical-Transition Broadening of a Two-Level System in Diamond. *ACS Photonics* **2023**, *10*, 2481-2487.

# Cryogenic Thermal Shock Effects on Optical Properties of Quantum Emitters in Hexagonal Boron Nitride

Thi Ngoc Anh Mai,[†] Sajid Ali,[‡] Md Shakhawath Hossain,[†] Chaohao Chen,[§,∥] Lei Ding, [#] Yongliang Chen,[††] Alexander S. Solntsev, [‡‡] Hongwei Mou,[#] Xiaoxue Xu,[#] Nikhil Medhekar[‡] and Toan Trong Tran[†,*]

[†]School of Electrical and Data Engineering, University of Technology Sydney, Ultimo, NSW, 2007, Australia

[‡]School of Physics and Astronomy, Monash University, Victoria 3800, Australia

[§]Department of Electronic Materials Engineering, Research School of Physics, The Australian National University, Canberra, Australian Capital Territory 2601, Australia.

[∥]ARC Centre of Excellence for Transformative Meta-Optical Systems (TMOS), Research School of Physics, The Australian National University, Canberra, Australian Capital Territory 2601, Australia.

[#]School of Biomedical Engineering, University of Technology Sydney, Ultimo, NSW, 2007, Australia.

[††]Department of Physics, The University of Hong Kong, Pokfulam, Hong Kong, China.

[‡‡]School of Mathematical and Physical Sciences, University of Technology Sydney, Ultimo, NSW, 2007, Australia.

*Corresponding author: trongtoan.tran@uts.edu.au

The Supporting Information includes:

**Figure S1** Spectra taken from 46 quantum emitters in hexagonal boron nitride before shock-cooling.

**Figure S2** A Lorentzian fit (blue line) to an exemplary spectrum (red circle) taken from a quantum emitter in hBN.

**Figure S3** Spectra taken from 40 quantum emitters in hexagonal boron nitride before slow-cooling.

**Figure S4** TEM thickness mapping of one representative hexagonal boron nitride flake before and after the cryogenic thermal shock cooling process.

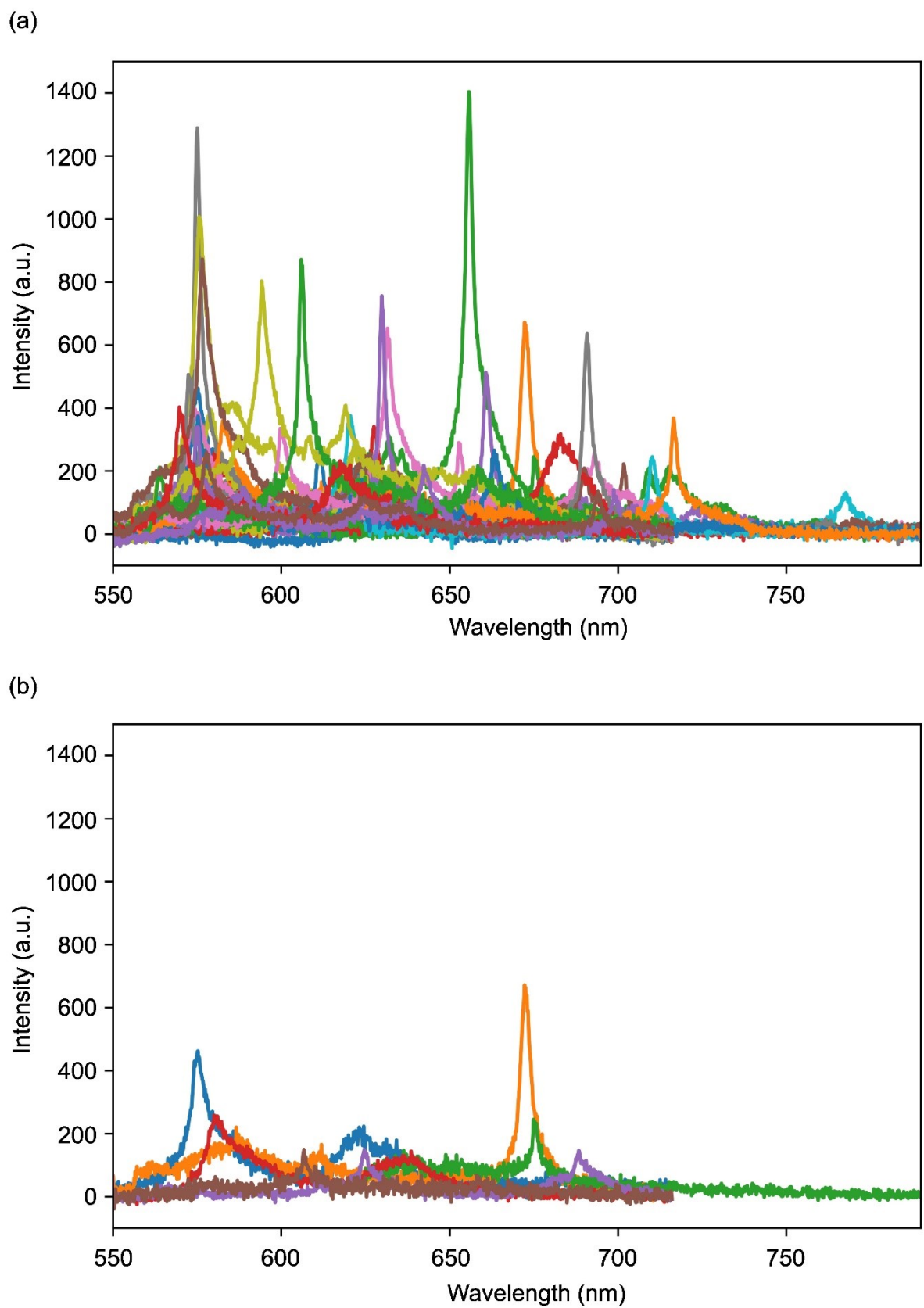

**Figure S1** a) Spectra taken from 46 quantum emitters in hBN before cryogenic thermal shock exposure. b) Spectra taken from the 6 emitters that bleached out after the thermal shock process.

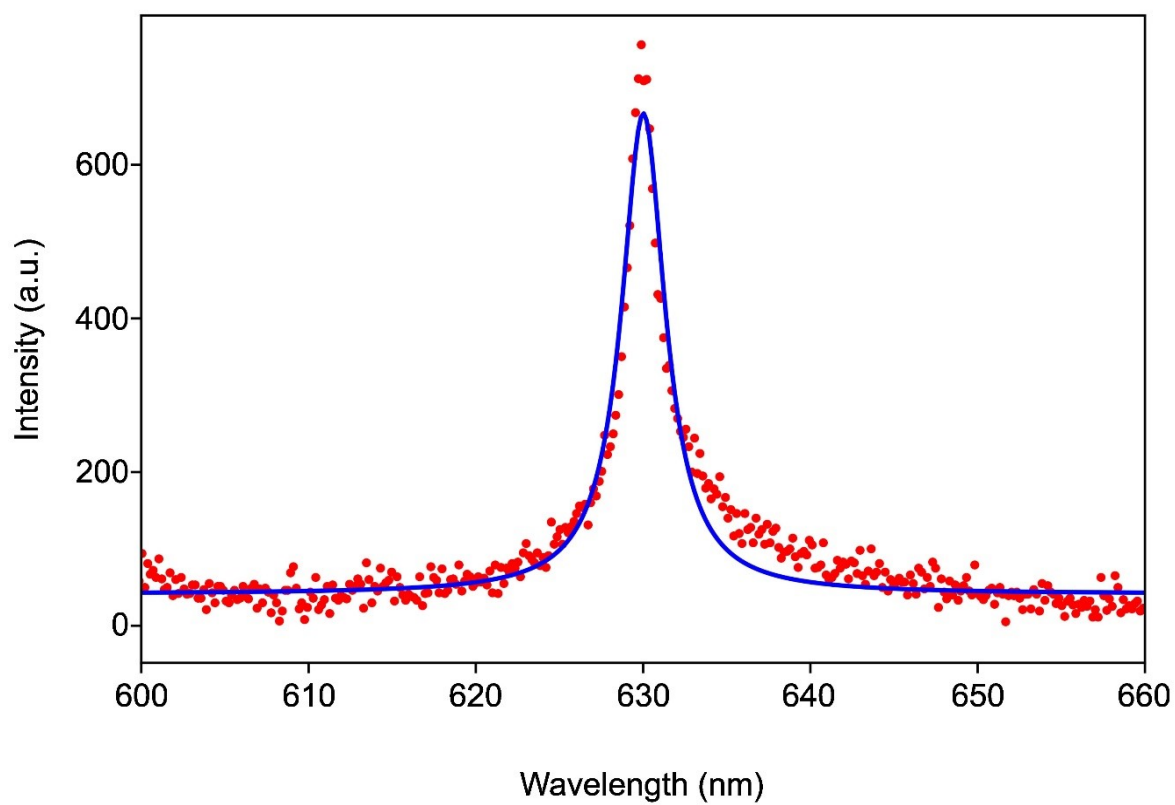

**Figure S2** A Lorentzian fit (blue line) to an exemplary spectrum (red circle) taken from a quantum emitter in hBN.

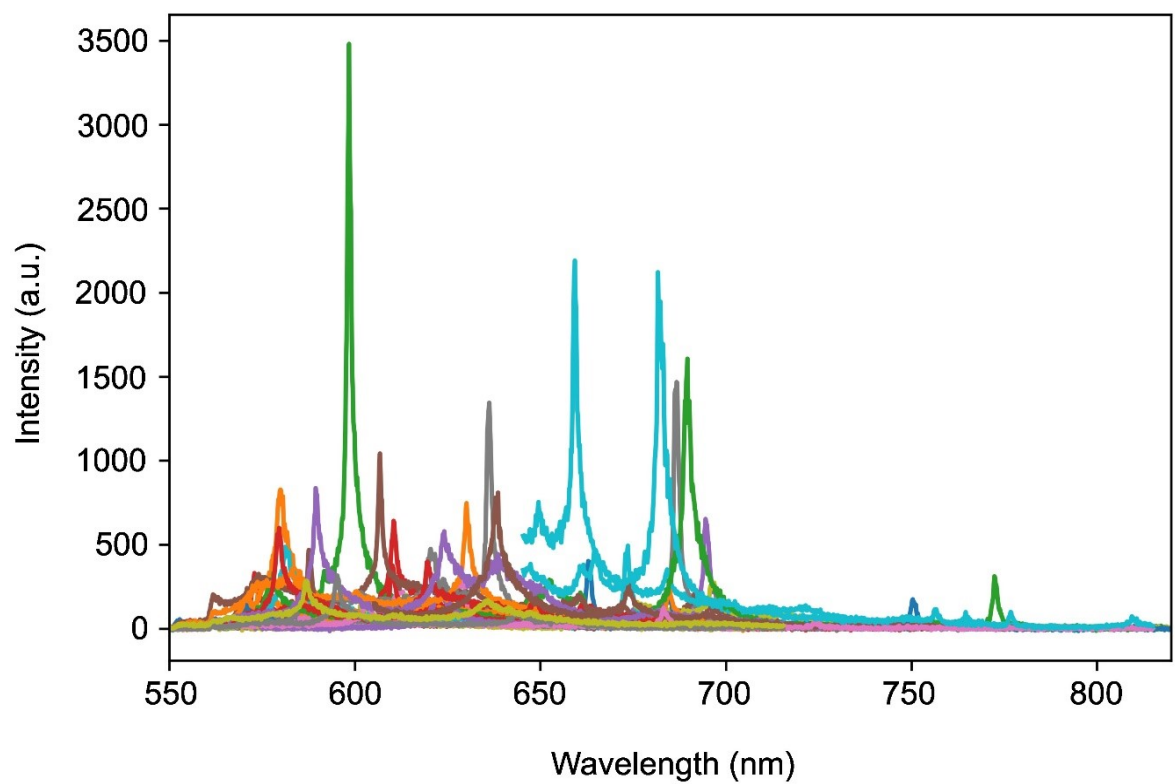

**Figure S3** Spectra taken from 40 quantum emitters in hBN before slow-cooling.

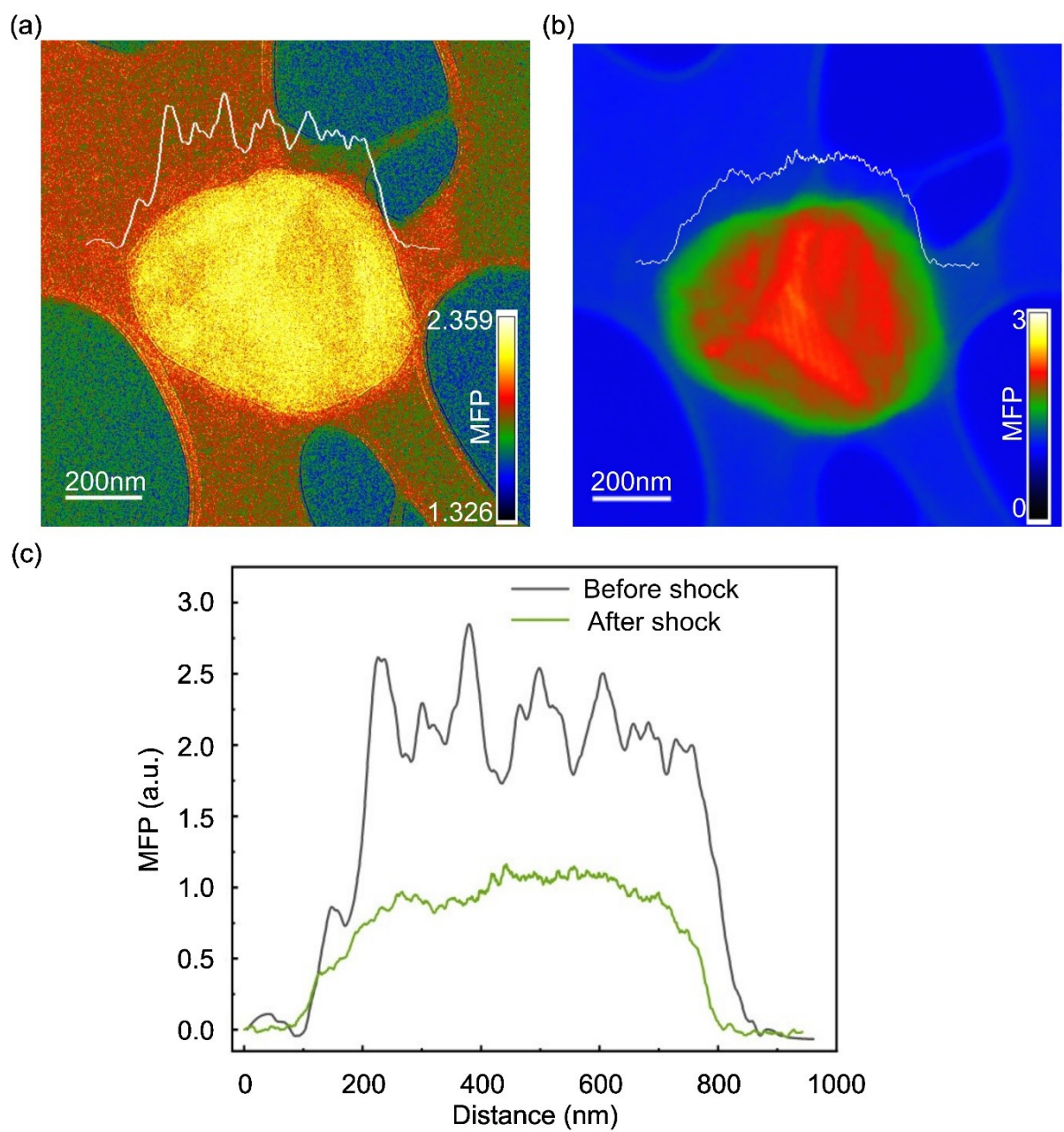

**Figure S4** (a-b) Transmission electron microscopy (TEM) thickness mapping of one representative hexagonal boron nitride flake before and after the cryogenic thermal shock cooling process, respectively. (c) The mean free path (MFP) line profiles of the hBN flake before and after the shock cooling exposure.